\documentclass{article}
\usepackage{a4}
\usepackage{epsfig}
\usepackage{amssymb}
\usepackage{amsmath}
\usepackage{rotate}

\begin{document}

\thispagestyle{empty}

\begin{flushright}
\today
\end{flushright}
\vspace*{1cm}
\begin{center}
{\Large{\bf A new experiment to search for the invisible decay of the orthopositronium\footnote{Based on
a talk given at Workshop on Positronium Physics, Zurich, Switzerland, 30-31 May 2003.}}}\\ 

\vspace{1cm}
{\large A.~Badertscher$^{1}$ , P.~Crivelli$^{1}$ , W.~Fetscher$^{1}$ , S.N.~Gninenko$^{2}$ , J.P.~Peigneux$^{3}$ , A.~Rubbia$^{1}$ , D.~Sillou$^{3}$ }

(1) Institut f\"{u}r Teilchenphysik, ETHZ, CH-8093 Z\"{u}rich, Switzerland \\
(2) Institute for  Nuclear Research, INR Moscow, Russia\\
(3) CNRS-IN2P3, France
\end{center}
\vspace{1.cm}
\begin{abstract}
\noindent
We propose an experiment to search for invisible decays of orthopositronium
(o-Ps) with a 90\% confidence sensitivity in the branching ratio 
as low as $10^{-8}$.   
Evidence for this decay mode 
 would unambigously signal new physics: either the existence of extra--dimensions or fractionally charged particles or new light gauge bosons.
The experimental approach and  the detector components  of the
proposed experiment are described.
\end{abstract}
\vspace{1cm}

\newpage
\pagestyle{plain} 
\setcounter{page}{1}
\setcounter{footnote}{0}

\title{\boldmath NEW PHYSICS AND THE INVISIBLE DECAY OF ORTHOPOSITRONIUM}


\section{Motivation}
 The Standard Model (SM) is very succesful, but is not yet a complete theory. There are several fundamental questions, which require physics 
beyond the SM for a solution. Among them are e.g. the hierarchy problem, the origin of the charge quantisation and the dark matter composition.
Although the new physics is expected typically at high energies, some models predict phenomena at low energies that might be observed in rare decays of the positronium, e.g. into a photon and an axion, a new particle
introduced for the solution of the strong CP problem, for a review see\cite{extradim,extradim2}. 

Positronium, the positron-electron bound state, 
 is the lightest known atom, which is bound and self-annihilates 
through the same, electromagnetic interaction. At the 
current level of experimental and theoretical precision this is  
the only interaction present in this system, see e.g. \cite{extradim,extradim2}. 
This feature has made positronium an ideal system for  
testing the accuracy of the QED calculations 
for bound states, in particular for the triplet ($1^3S_1$)
state of $Ps$, orthopositronium ($o-Ps$). 
Due to the odd-parity under
C-transformation  $o-Ps$ decays
predominantly into three photons. 
 As compared to the singlet ($1^1S_0$) state (parapositronium),
 the long lifetime of $o-Ps$ decay, (due to the phase-space and  additional
$\alpha$ suppression factors) gives an enhancement factor $\simeq 10^3$,
in the sensitivity to an admixture of 
new interactions, which are not accommodated in the SM.

This paper is focused on the search for  $o-Ps \rightarrow invisible$ decay, i.e. a process which is not accompanied by energy deposition in a detector. Within the SM orthopositronium  can  decay invisibly into a
 neutrino-antineutrino pair.  
The $o-Ps \to \nu_e \bar{\nu}_e$ decay occurs through 
$W$ exchange in the $t$--channel and $e^+e^-$ annihilation via $Z$.
The decay width is \cite{czar}
\begin{eqnarray} 
\Gamma(o-Ps \to \nu_e  \bar{\nu}_e) 
\approx 6.2 \cdot 10^{-18}\Gamma(o-Ps \to 3\gamma)
\end{eqnarray}
For other neutrino flavours only the $Z$-diagram 
contributes. For $ l \neq e$ the decay width is \cite{czar}
\begin{eqnarray} 
\Gamma(o-Ps \rightarrow \nu_l \bar{\nu}_l) 
\approx 9.5 \cdot 10^{-21}\Gamma(o-Ps \to 3\gamma)
\end{eqnarray}
Thus, in the SM the $o-Ps \to \nu \bar{\nu}$ decay rate
is very small and evidence for invisible decays  
 would unambiguously signal the presence of new physics.  
  It may be worthwhile to remember that the process with analogous experimental signature, 
$Z \rightarrow invisible$ decay 
plays a fundamental role in the determination of the number of lepton families.  

The new models that are relevant to the $o-Ps\to invisible$ decay mode 
predict the existence either of i) extra-dimensions \cite{extradim2,tinyakov}, or ii) fractionally charged particles\cite{holdom}, or iii) a new light vector gauge boson\cite{extradim}, or
 iv) dark matter of the mirror matter type\cite{glashow,foot}. 
To test these models, the required sensitivity in the 
branching ratio $Br(o-Ps\to invisible)$  
has to be at least as low as $10^{-8}$. 
Note however that the search for the mirror matter  requires  in addition
the experiment to be performed in vacuum
(for more details see \cite{foot}).
Such an experiment  is discussed and proposed in \cite{Badertscher:2003rk}.

The first experiment to search for invisible decay channels of o--Ps 
was performed a long time ago \cite{ajotan}. The best present limit for the branching ratio, set in 1992 by Mitsui et al.\cite{mits} is
\begin{equation}
 Br(o-Ps \rightarrow invisible ) < 2.8 \times 10^{-6}
\end{equation}

The paper is organized as follows.
In the next sections the physics of items i)--iii) is briefly discussed.  The experiment, its 
setup components and the expected sensitivity are discussed in Sections 2, 3 and 4 
respectively. Section 5 contains concluding remarks.

\subsection{Extra dimensions}

Models with infinite additional dimensions of the Randall-Sundrum type (brane-world models) with a big compactification 
radius \cite{randall}${}--$\cite{rubilnik} could provide the natural solution to the gauge hierarchy problem.

Recently Dubovskj, Rubakov and Tinyakov \cite{tinyakov} pointed out that in R--S brane world models with localized bosons, massive gauge bosons localized to the Plank brane are always unstable. The reason is that bulk modes can have arbitrarily small masses. Any massive mode localized to the brane is then kinematically allowed to decay into these modes.\\
In a recent paper \cite{extradim2} this mechanism was applied to calculate the decay of o--Ps into invisible bulk modes and obtained
\begin{eqnarray}
Br(o-Ps \rightarrow \gamma^{*} \rightarrow additional~dimension(s))& 
=\\ \nonumber 
\frac{9\pi}{4(\pi^2-9)}\cdot \frac{1}{\alpha^2}\cdot \frac{\pi}{16}
(\frac{m_{o-Ps}}{k})^2   \approx 3\cdot10^{4}(\frac{m_{o-Ps}}{k})^2
\end{eqnarray}
For a solution of the gauge hierarchy problem the parameter $k$ is 
expected to be $k \le O(10)~TeV$.  Indirect measurements at  LEP of
the  $\Gamma(Z \to invisible)$ decay width constrain the parameter  to  $k \ge 2.7~TeV$.  Using this bound one can find  
\begin{equation}
Br(o-Ps \rightarrow additional~dimension(s)) \le 0.4 \cdot 10^{-9}
\end{equation}
 
The direct LEP measurements of $\Gamma(Z \to invisible)$
 results in a less stringent limit \cite{extradim,extradim2}.
\begin{equation}
Br(o-Ps \rightarrow additional~dimension(s)) \le 8\cdot 10^{-8}
\end{equation}
These estimates  show that the region of $\simeq  10^{-8}- 10^{-9}$ for the branching ratio is of great interest for the observation of extra--dimensions.
 
\subsection{Millicharged particles}

In 1986 Holdom \cite{holdom} showed that
grand unified models can be constructed in a natural way, adding a second, 
unobserved, photon to the interaction Lagrangian. 
In this type of model particles with an electric charge very small compared to the electron, are predicted. If such  milli-charged particles exist with a small mass, the $o-Ps$ could decay apparently invisibly, since the particles would most likely penetrate any type of calorimeter  without interaction.
The corresponding decay width is \cite{ignatiev}

\begin{equation}
\Gamma(o-Ps \rightarrow X \bar{X}) = \frac{\alpha ^5 Q^2_{X}m_e}{6}\cdot 
k \cdot F(\frac{m^2_X}{m^2_e}),
\end{equation}
where $Q_{X}$ is the electric charge of the $X$-particle $(Q_e \equiv 1)$,
 $k = 1$, 
$F(x) = (1 - \frac{1}{2} x)(1-x)^{\frac{1}{2}}$ for spin $\frac{1}{2}$ and
  $k = \frac{1}{4}$,
$F(x) = (1-x)^{\frac{3}{2}}$ for a spin-less $X$-particle. 
For spin $\frac{1}{2}$ millicharged $X$-particles and $m_X \ll m_e$ one 
can find from the experimental bound (3) 
that  $Q_{X} \le 8.6 \cdot 10^{-5}$  \cite{mits}.A search for 
millicharged particles through the $o-Ps \to invisible$ decay 
with the sensitivity of
$Br(o-Ps \to invisible)\simeq 10^{-8}$ would touch  the parameter space 
which is not excluded by results of the recent direct experiment at 
SLAC \cite{prinz}.

\subsection{New light $X$-boson}

A light vector $X$-boson with 
the interaction Lagrangian
\begin{equation}
L_X =g_X \bar{\psi} \gamma_{\mu} \psi X^{\mu}
\end{equation}

 will also lead to invisible decays of 
$o-Ps$.\footnote{For the recent  phenomenological bounds in models with a 
light vector $X$-boson related to the muon $(g-2)$ and 
so-called NuTeV anomalies see, respectively \cite{gnkr}, \cite{david} 
and \cite{dob}.} Supposing that the $X$-boson interacts with 
other particles (fermions) or (as a consequence of the Higgs mechanism) 
with itself and a scalar the contribution of the $X$-boson to the electron anomalous 
magnetic moment is given by the well known formula
\begin{equation}
\delta a_e = \frac{\alpha_{X}}{\pi}\int^{1}_{0}\frac{x^2(1-x)}{x^2 +(1-x)
\frac{m^2_X}{m^2_e}}
\end{equation}
For $m_X \ll m_e$ from the bound arising from the determination of the fine structure constant with the Quantum Hall Effect (QHE)\cite{extradim}, one can find that  
$\alpha_{X} < 3 \times 10^{-10}$. For the case 
of the heavy $X$-boson ($m_X \gg m_e$) the bound on the anomalous 
electron magnetic moment leads to the bound
\begin{equation}
\alpha_{X}\frac{m^2_e}{m^2_X} < 4.5 \cdot 10^{-10}
\end{equation}
For the reaction $o-Ps \rightarrow X^{*} \rightarrow X_1 \bar{X}_1$ 
(here $X^{*}$ is a virtual $X$-boson and $X_1$ is a fermion 
(sterile neutrino) or a scalar particle) one finds 
that
\begin{eqnarray}
Br(o-Ps \rightarrow X^{*} \rightarrow X_{1}\bar{X}_{1}) 
= \\ \nonumber
\frac{3\pi}{4(\pi^2 - 9)}
\cdot k \cdot F(\frac{m^2_{X_1}}{m^2_e})   
(1 -\frac{m^2_{X}}{m^2_{0-Ps}})^{-2}\frac{\alpha_{X} \alpha_{XX_1}}
{\alpha^3},
\end{eqnarray}
where $F(x)$ has been defined before and 
$\alpha_{XX_1} = g^2_{XX_1}/4\pi$.  From the bound of the QHE we find 
for $m_X \ll m_e$ that 
\begin{equation}   
 Br(o-Ps \rightarrow X^{*} \rightarrow X_{1}\bar{X}_{1}) \le
k \times 2 \cdot 10^{-3}\cdot \alpha_{XX_1}
\end{equation}
To have an experimentally interesting branching ratio of the order 
$O(10^{-8})$, setting  $\alpha_{X} = 3 \times 10^{-10}$ we must 
have $\alpha_{XX_1} \sim
5 \cdot 10^{-6}$.  

In the opposite limit $m_X \gg m_e$ the corresponding bound reads
\begin{equation}   
 Br(o-Ps \rightarrow X^{*} \rightarrow X_{1}\bar{X}_{1}) \leq
k \times 3 \cdot 10^{-3}\cdot \alpha_{XX_1} \cdot \frac{m^2_e}{m^2_X}
\end{equation}
For an $X$-boson mass close to 
the orthopositronium mass, we have the enhancement factor 
$(1 - \frac{m^2_X}{m^2_{o-Ps}})^{-2}$ in formula (11) for the branching 
 ratio, and hence, the coupling constant $\alpha_{X}\cdot\alpha_{XX_1}$ could be 
smaller.

\section{Proposed setup and experimental technique}

\begin{figure}{\label{front}}
\centerline{\psfig{file=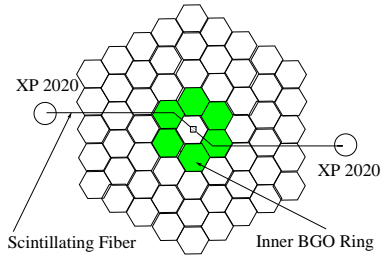,width=5cm}}
\vspace*{8pt}
\caption{Front view of the calorimeter.}
\end{figure}

\begin{figure}{\label{side}}
\centerline{\psfig{file=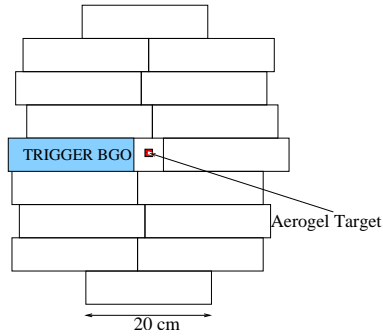,width=5cm}}
\vspace*{8pt}
\caption{Side view of the calorimeter.}
\end{figure}

In this section the design of the experiment to search for $o-Ps \to invisible$ with a sensitivity better than $10^{-8}$ is presented.
The main components of the detector (shown in Figs. 1, 2) are: 
the positron source (${}^{22}$Na), the positron tagging system, composed of a scintillating fiber viewed by two photomultipliers (PM), the positronium formation $SiO_2$  target and a  hermetic $\gamma$--detector.
 The coincidence of the PM signals (see Fig.\ref{front}) from the positrons crossing the fiber, opens the gate for the data acquisition (DAQ). In the off--line analysis the 1.27 MeV photon, which is emitted from the source simultaneously with the positron, is required to be in the trigger BGO counter (shown in Fig.\ref{side}) resulting in a high confidence level of positron appearence in the positronium formation region. A positron, which enters the $SiO_2$ target may capture an electron creating positronium. The calorimeter detects, either the direct $2\gamma$ annihilation in flight or the 2(3) photons from the para (ortho)--positronium decays in the target.  

The occurrence of the $o-Ps \to invisible$ decay would appear as an excess of events with zero--energy deposition in the calorimeter above those expected from the Monte Carlo prediction (see section 4) or from the direct background measurement. 

This measurement presents a new feature of this type of experiment. The idea is to obtain a pure o--Ps decay energy spectrum by comparing two different spectra from the same target filled either with $N_2$ (low o-- Ps quenching) or with air, where the presence of  paramagnetic $O_2$ will quench the fraction of o--Ps in the target from 10$\%$ down to 3$\%$, due to the spin exchange mechanism\cite{kalimoto}: 
\begin{equation}
o-Ps+O( \uparrow \uparrow) \to p-Ps + O( \downarrow \uparrow)
\end{equation}
 Thus the subtraction of these properly normalized spectra will result in a pure o--Ps annihilation energy spectrum in the $\gamma$--detector.

In addition, compared to the previous experiments\cite{ajotan,mits}, the region around the target has been designed with as little dead material as possible in order to reduce the photon absorption. The confidence level for tagging the positron appearance in the target and the efficiency of annihilation photon detection have been improved.

\subsection{Source of positrons}
The radioactive isotope ${}^{22}$Na, which has a half--life of 2.6 years, was chosen as a source of positrons. The positron emission from the source is accompanied by the prompt emission of a 1.27 MeV photon ($\simeq$ 3 ps delay) (see Fig.3). This signature provides an inefficiency for positron tagging less than $\simeq 10^{-9}$.
The source activity was selected based on a compromise between required trigger rate and a decrease in the signal efficiency due to the overlap of events close in time (pile up effect). The optimal source activity has been determinated to be about 50 kBq, for which, according to the simulation, the signal efficiency drops down to $84\%$ (the signal of $o-Ps \to invisible$ is defined as an event depositing  less than 50 keV in the $\gamma$--detector). In $10\%$ of the cases the 1.27 MeV photon from the source is emitted due to the electron capture (EC), without positron emission. As discussed in the next section, this process introduces background related to the source container material, which was eliminated by depositing the radioactive isotope directly on the scintillating fiber. A comparison between ${}^{22}$Na and other $\beta ^+$ radioactive isotopes with a sufficiently long lifetime has been made to combine the required confidence level for positron tagging with the minimal amount of dead target material and a high stopping efficiency of positrons in the target. Taking all these considerations into account, we conclude that the ${}^{22}$Na is the best compromise for our measurements.

\begin{figure}\label{22Na}
\centerline{\psfig{file=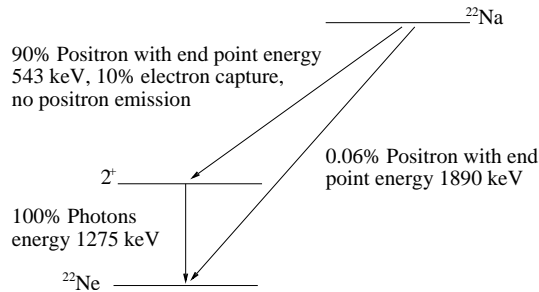,width=8cm}}
\vspace*{8pt}
\caption{Decay scheme of ${}^{22}$Na }
\end{figure}

\subsection{The positron tagging system}
The tagging system for the experiment was designed in order to have a confidence level for the positron appearance better than $10^{-9}$. It is composed of a thin scintillating fiber viewed by two PM's. It provides a relatively high efficiency and a small amount of dead material.  

\subsubsection{The scintillating fiber}
For a good positron tagging efficiency , the attenuation length and the light yield of the fiber are crucial parameters. After testing the response of different scintillating fibers, the Bicron BFC 12 with a diameter of 0.5 mm  has been selected. They have been prefered over the 1 mm fibers used in the experiment\cite{bader} because they have less material and the possibility to have a more hermetic calorimeter, due to the smaller gaps between the crystals, through which the fiber has to penetrate to transport the signal outside the calorimeter volume. Similar to experiment\cite{bader} the fiber is squeezed in the middle to get a thickness of about 100 $\mu$m. The ${}^{22}$Na source is deposited in the middle of the squeezed region. The measured efficiency for positron detection is slightly less than 50$\%$.

\subsubsection{The photomultipliers}
The PM's XP 2020 have been chosen for their low noise and for their spectral sensitivity, that fits the emission spectrum of the fiber. The use of two PM signals in coincidence decreases the probability of accidental coincidences with the 1.27 MeV photon to a level of $3.2\times 10^{-11}$ (see section 3). In the future we plan to test Hamamatsu H3165-11 PM's with base R647, because of their high efficiency for single photon counting.\footnote{We would like to thank Dr.S.Asai for his advice\cite{mits}.} 

\subsection{Positronium production target}
 A $SiO_2$ aerogel hemisphere with a radius of 5 mm and a density of 0.1 $g/cm^3$, where positronium can be formed with about 
 5$\%$ probability (as measured directly from the lifetime spectrum), is used as a target. The aerogel is a porous material, where the o--Ps migrating in the intergranular space can decay almost freely. The dimension has been optimised to minimize the dead material, while keeping,  enough material to stop positrons efficiently. The simulation shows that the probability that more than 900 keV photons energy is absorbed in it, for the given dimension, is less than $10^{-10}$. Before its installation the aerogel is evacuated in a vacuum chamber (10$^{-3}$ Torr) and backed out for 2 hours to remove the water from the pores. Then the target is filled either with high purity Nitrogen or with air. 


\subsection{Calorimeter}

The optimal choice of the $\gamma$--detector (ECAL) can be made by the following considerations. The total ECAL mass W is given roughly by 
\begin{equation}
W \simeq 4\pi/3 \rho L^3
\end{equation}

where $\rho$ and $L$ are respectively mass density and the radius of the ECAL
detector. We chose $L\simeq 20\lambda_{511}$, where $\lambda_{511}$
is the attenuation length of 511 keV $\gamma$'s. The relevant parameters for different types of materials used in ECAL's are listed below in Table 1.

\begin{table}[h]
\begin{center}
\begin{tabular}{|c|c|c|c|c|}
\hline
 ECAL type   & BGO&NaI&CsI(Tl)&Sc plastic/liquid   \\
                   & &&&  \\
\hline
\hline
                   & &&&  \\
Attenuation length, at 511 keV &  $\simeq$1 cm &  $\simeq$2.5 cm &  $\simeq$1.9 cm & $\simeq$10 cm  \\
                   & &&&  \\
\hline
                   & &&&  \\
density,  $g/cm^3$       & 7.1 & 3.6&4.5&   1.0 \\
                   & &&&  \\
\hline
                   & &&&  \\
Required ECAL mass, kg     & $\simeq$240&  $\simeq$1890& $\simeq$1034& $\simeq$33510\\
                   & &&&  \\
\hline
                   & &&&  \\
Light yield, $N_{\gamma}/511$keV & $\simeq 4\cdot10^3$ &$\simeq 20\cdot 10^3$
& $\simeq 10\cdot 10^3$ & $\simeq 2\cdot10^3$\\
                   & &&&  \\
\hline
                   & &&&  \\
Hygroscopic                   & no & yes & yes/slightly& no  \\
                   & &&&  \\
\hline
\hline
\end{tabular}
\end{center}
\caption{Comparison between different types of ECAL}
\end{table}

The required mass is minimal for a BGO ECAL due to its high effective $Z$ (remember that the photo-absorption cross-section $\sigma\sim Z^5$). 
Another important feature of BGO's is that they are not hygroscopic, thus, no additional dead material has to be introduced.
For the crystal wrapped in aluminized mylar the light yield was measured to be 200$\pm$14 photoelectrons/1 MeV. This results in a probability of zero energy detection due to Poisson fluctuation of the number of photoelectrons, to be less than $10^{-11}$ for the zero energy signal defined as events with energy deposition less than 50 keV\cite{bader}.
A disadvantage in using BGO crystals is the ~300 ns decay time of 
their scintillations, 
which may cause a drop of efficiency for the signal events due to the
pile up effect. However for the source intensity planned to be used
in the experiment this drop of efficiency is calculated to
be about 15\%. 
 Thus, these results justify the selection of BGO as the $\gamma$--detector.
The crystals, which have been lent to us by the Paul Scherrer Institute (Villigen, Switzerland), have a hexagonal shape with a length of 20 cm and an outer diameter of 5.5 cm, their original wrapping is a 0.75 mm thick teflon. In order to reduce this amount of dead material, the inner ring of BGO's has been wrapped in a 2 $\mu$m thick foil, aluminized from both side with 1000 $\AA$ thick layers. The required number of crystals ($\simeq 100$), determined with the simulations, provides an almost isotropically uniform thickness of 20--22 cm of BGO.

\subsection{Monitoring}
Every crystal is equipped with a LED, fired with a pulser, to monitor the crystals response. The gas and light tight box is thermally stabilised to $\pm 1$${}^o$C. The HV for every PM is measured periodically during the run. 

\subsection{Data Acquisition}
The data acquisition is started by the gate generated from the coincidence between the two XP 2020 of the scintillating fiber. The gate has a length of 3.3 $\mu$s in order to suppress the probability of o--Ps decaying after this time to the level of $10^{-11}$. The energy of each crystal and  the two XP 2020 signals from the fibers are recorded with four 32 channels CAEN QDC's v792. The timing between the signal from the fiber and the crystals of the endcaps and the inner most ring of the calorimeter is measured by a LeCroy 1182 TDC. The VME crate is interfaced to a PC with a National instruments VXI-MXI2 system and the data are acquired with LabView. The monitoring is done by another computer using ROOT, not to slow down the data acquisition.

\section{Background}
The main background sources, summarized in Table 2, can be divided into three different types:
\begin{itemize}
\item A real trigger accompanied by annihilation energy losses.
	\begin{itemize}
        \item Hermiticity: as pointed out in the previous section (Table.1), the 511 keV photons have an attenuation lenght of about 1cm in BGO. For 20 cm BGO the escape probability is about $10^{-9}$.  
        \item Absorption of photons in dead material: some energy of the photons can be absorbed either in the target or/and in the wrapping of the crystals or/and in the fiber.
        \item Absorption of photons in the trigger energy window: the resolution of the endcap at 1.27 MeV is about 16$\%$ FWHM. If one (or more if e.g. one photon is backscattered in the target) of the positronium decay or direct annihilations photons overlap with the 1.27 MeV in a crystal it can be absorbed in the trigger energy window due to the resolution of the crystals.
	\end{itemize}
\item A fake trigger, with no positron in the detector, can be produced by the EC 1.27 MeV photon. This gives two sources of background:
	\begin{itemize}
        \item the 1.27 MeV photon can scatter with an electron in the fiber and the electron can deposit some fraction of its energy in the fiber and opens the gate for the DAQ. In this case the background level is about $10^{-6}$. To reduce it, a trigger requires the coincidence between the signal from the fiber and the signal from the opposite (to the direction of the positron entering the fiber and the aerogel) endcap counter. This reduces the background by one order of magnitude. The next step is to suppress the background related to the 1.27 MeV photon scattering in the material, which surrounds the source,by depositing the $^{22}$Na directly on the fiber. This lowers the background by another order of magnitude. The remaining background is associated with the 1.27 MeV EC photon scattering back to the trigger BGO (this photon will have between 200--300 keV energy) and a Compton electron multiple scattering in the fiber hitting the same trigger counter. This seems to be the most dangerous background for our experiment. A possible solution, which is under study, will be discussed in Section 6.  
        \item The other possibility is that this photon may accidentally coincide with a fake positron trigger generated by the PM noise.
	\end{itemize}

\item Physical backgrounds: 

     \begin{itemize}
\item single photon or photonless decays of o--Ps ions: $o-Ps^{-}\to e^- +\gamma$, $o-Ps^{--}\to e^- +e^- $
\item photonuclear absorption of annihilation photons accompanied either by photo--neutrons or by excitation of long lived nuclear states.
     \end{itemize}

The detailed evaluation of this background is in progress; some preliminary estimates show that it is less than $10^{-10}$. The o--Ps produced in an excited state, immediately de--excites to the ground state\cite{gidley} due to the high collisional ($10^4$ collision/lifetime) quenching in the $SiO_2$ . Since the experiment is not performed in vacuum the positron wave function is assumed to overlap always with an electron wavefunction. The largest lifetime of the positrons in our experiment is assumed to be those for the three photon decay of o--Ps, namely 142 ns, thus making the probability of the positron not to annihilate during the 3.3 $\mu$s gate negligible.
\end{itemize}

\section{Monte--Carlo}
The feasibility of the experiment has been studied with a  Monte Carlo simulation based on the code Geant 3. The simulation was cross--checked and tuned with our previous search for an exotic three--body decay of o--Ps \cite{bader}. The inefficieny for 511 keV $\gamma$ detection in the measurements was predicted to be $(1.0\pm 0.2)\times 10^{-4}$. The comparison with the data shows an agreement within $30\%$. The low energy part of the spectra, shown in Fig.4, is quite well reproduced, nevertheless more work has still to be done. The losses due to light transportation, self absorption and also the non uniformity of the crystals have to be included.\\

\begin{figure}\label{etot_comp}
\centerline{\psfig{file=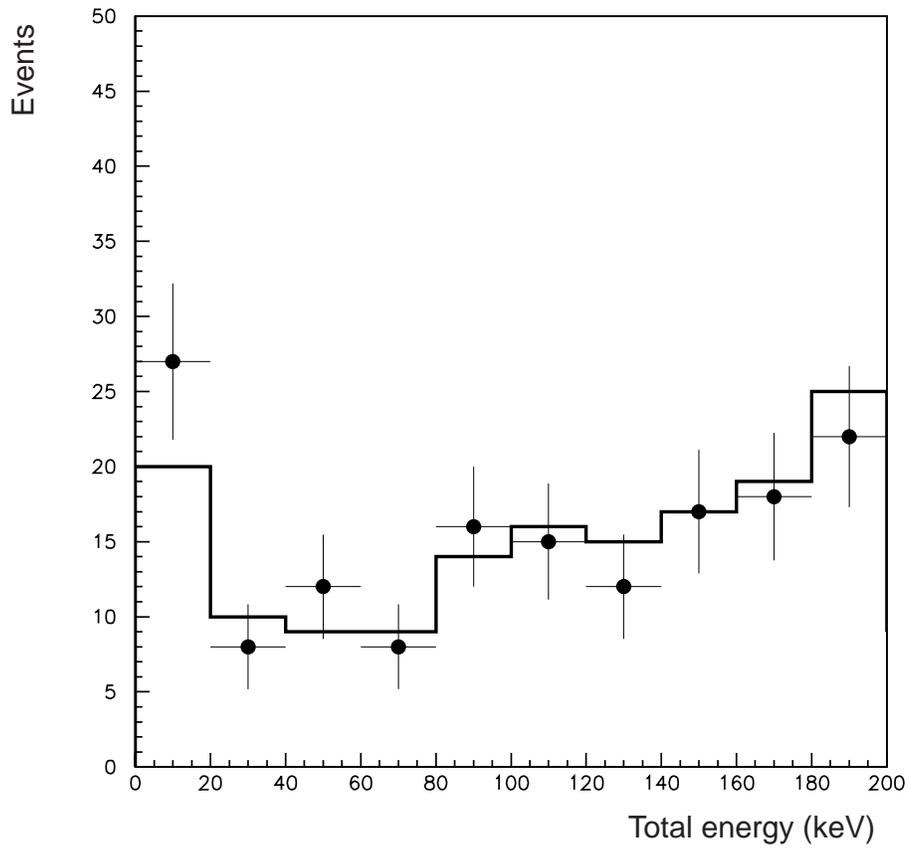,width=12cm}}
\caption{Comparison between MC simulation (points) and data of the experiment (dashed line)}
\end{figure}

\begin{table}[h]
\begin{center}
\begin{tabular}{|c|c|}
\hline
Background source	&   Expected level\\
\hline
\hline
Hermiticity & $< 10^{-10}$ \\
Dead Material & $2\times 10^{-10}$ \\
Absorption in trigger energy window & $10^{-9}$ \\
\hline
Compton EC photon & $1.2\times 10^{-8}$  \\
Accidental noise and EC photon & $3.2\times 10^{-11}$ \\
\hline
Physical backgrounds & $10^{-10}$\\
\hline
\hline
\end{tabular}
\end{center}
\caption{Expected background level}
\end{table}


\section{Sensitivity}

As mentioned before, the trigger is given by the coincidence between the signals from the two PM of the fiber. Off--line, the 1.27 MeV photon is selected in the trigger counter inside the energy window 1170--1470 keV (see Fig. 5). In order to reduce the accidental background (see previous section) the time between the coincidence and the 1.27 MeV $\gamma$ is required to be inside the time window as shown in Fig. 5. A cut on the energy of the PM is applied to reduce the accidental noise. The distance between the source and the trigger BGO has been choosen to be 2 cm as a compromise between the trigger rate and the background from the Compton EC photon. The expected trigger rate for a source intensity of 50 kBq is
\begin{equation}
Trigger~rate = \frac{1}{2}\cdot Positrons~emission~rate\cdot 5\% \cdot Fiber~efficiency
\end{equation}
and knowing the fraction of o--Ps in the spectra, which is about $5\%$
\begin{equation}
Number~of~o-Ps/day = Trigger~rate \cdot o-Ps~fraction \cdot 86400 s\simeq 3\times 10^{6}
\end{equation}
Using Poisson statistics and assuming no background event observed, one can calculate the sensitivity of the experiment\cite{particle} after 60 days of running:
\begin{equation}
Br(o-Ps \rightarrow invisible) = 2.3/1.8\times 10^{8} \simeq 1.3\times 10^{-8}
\end{equation}

\begin{figure}\label{cuts}
\centerline{\psfig{file=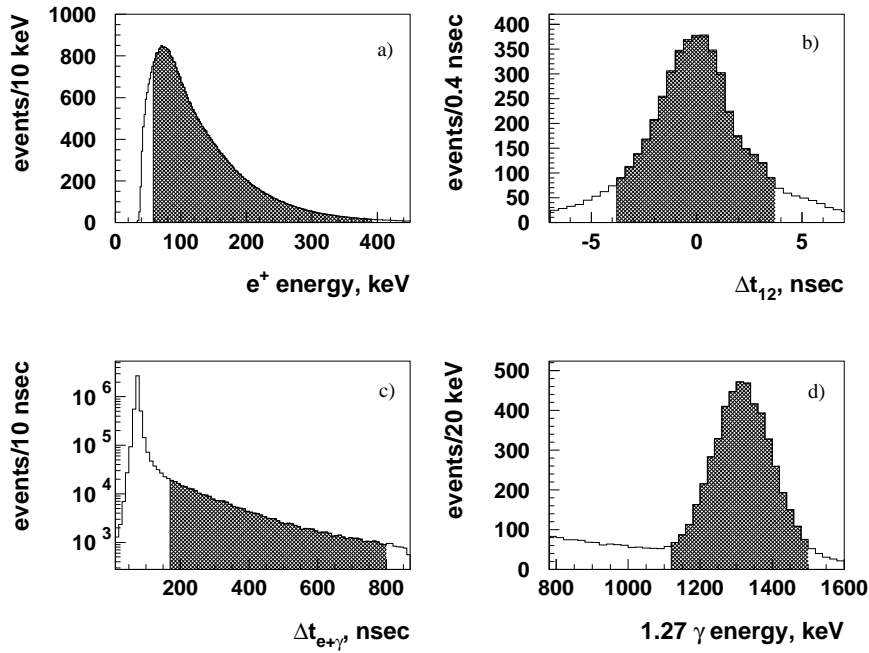,width=12cm}}
\caption{Cuts for the event selection: a) Cut on the energy deposited in the scintillating fiber, b) Cut for the time difference between the two XP2020 from the fiber, c) Time spectra between start from the coincidence between the two XP and the stop from the trigger BGO, d) Energy window for 1.27 MeV photon in the trigger BGO. The shaded area corresponds to accepted events.}
\end{figure}

\section{Conclusions}

With the presented type of calorimeter a background free limit for the $Br(o-Ps\to invisible)$  of $\simeq 10^{-8}$ is reacheable in about two months of data taking.
Although not all the possible background sources have been identified yet, the dominant background seems to be associated with the EC photon that scatters in the fiber back to the trigger BGO counter. The accompanying Compton electron can deposit $\simeq$ 50--100 keV in the fiber, which is sufficient to trigger the coincidence of the PM's. Since the energy deposition in the trigger counter is almost equal to the total initial energy of the 1.27 MeV photon, it is quite difficult to reject this type of events with the given resolution of the crystal. One possible way to discriminate these $e^- \gamma$ pairs is to use a thin plastic scintillator in front of the trigger BGO, with a faster timing response than the one of the BGO's. This work is in progress.


\section*{Acknowledgements}

This experiment was mainly possible due to the availability of the crystals, which were lent to us by PSI, Villigen.
We are in debt to Jacques Wolf of the CERN radioactivity department for the source deposition on the fiber. 
We thank U. Gendotti and  V. Samoylenko for helping in the development of 
the efficient positron tagging system, N.Golubev for the help in the preparation of the $SiO_2$ target.
This work was supported in part by ETH/Zurich and the Swiss National Research Foundation.


\end{document}